\documentclass[5p]{elsarticle}

\usepackage{graphicx}
\usepackage{dcolumn}
\usepackage{array}   
\newcolumntype{L}{>{$}c<{$}} 
\usepackage{bm}
\usepackage[justification=raggedright]{caption}
\usepackage{subcaption}
\usepackage{svg}
\usepackage{physics}
\usepackage{booktabs}
\usepackage{algorithm}
\usepackage{algorithmic}
\usepackage{tabularx}
\usepackage{comment}
\usepackage{xcolor} 
\usepackage{hyperref}
\hypersetup{colorlinks = true,linkcolor = blue, breaklinks = true}

\usepackage[normalem]{ulem}

\journal{Nuclear Materials and Energy}

\begin{document}

\begin{frontmatter}
\title{Langmuir probe and infrared thermography measurements of wide and narrow heat flux profiles in the ST40 tokamak}
\author[TE]{T. Brewer}
\author[TE]{X. Zhang\corref{cor1}}
\ead{Laura.Zhang@tokamakenergy.com}
\author[TE]{C. Marsden}
\author[LK]{A. Scarabosio}
\author[TE]{J. Bryant}
\author[TE]{P. F. Buxton}
\author[TE]{M. Fontana}
\author[ORNL]{T. Gray}
\author[TE]{\\N. Lopez}
\author[TE]{L. Martinelli}
\author[TE]{A. Rengle}
\author[ORNL]{E.J.C. Tinacba}
\author[TE]{E. Vekshina}
\author[TE]{H. V. Willett}

\address[TE]{Tokamak Energy Ltd, Abingdon, United Kingdom}
\address[LK]{LINKS Foundation, 10138 Turin, Italy}
\address[ORNL]{Oak Ridge National Laboratory, TN, United States of America}
\cortext[cor1]{Author to whom correspondence should be addressed to}


\begin{abstract}
Very narrow heat flux profiles have been observed on ST40 with decay lengths in the near scrape-off-layer (SOL) of $\lambda_{q,near}<$~1~mm in H-mode plasmas. Measurements from divertor Langmuir probes are compared with an infrared (IR) thermography system for upper-single null diverted plasmas in L-mode and ELM-free H-mode. SOL current to the grounded divertor targets is measured from the Langmuir probes with profiles exhibiting a similar exponential decay to $\lambda_{q,near}$ suggesting the near SOL heat flux is related to enhanced electron current to the divertor. Inclusion of electron current contributions to the heat flux transmission coefficient is crucial in capturing similar narrow profiles to those from the IR thermography system. Ongoing upgrades to ST40 will enable more investigations on SOL power decay lengths on inboard and outboard targets in double and single null configurations.
\end{abstract}
\end{frontmatter}

\section{Introduction}
\label{sec:Intro}
Recent experimental campaigns conducted on the ST40 tokamak have investigated heat flux profiles in the upper-outer divertor region with a focus on heat flux decay lengths, $\lambda_q$, in the scrape-off-layer (SOL)~\cite{2024_Zhang_etal, 2024_Marsden_etal}. Namely these studies have focused on the observation of two decay length groupings, one  ``wide'' and one ``narrow''. The wide group is observed in ST40 L-mode discharges and follows H-mode scalings based on a multi-machine database~\cite{2013_Eich_etal} while the narrow group is observed in ST40 H-mode discharges with decay lengths $<$~1~mm. 

Narrow near SOL profiles have been observed for limited configuration dishcarges in COMPASS~\cite{2015_Horacek_etal}, DIII-D~\cite{2015_Stangeby_etal}, JET~\cite{Arnoux_2013} and TCV~\cite{2017_Nespoli_etal} and in diverted plasmas in ASDEX-U~\cite{2014_Carralero_etal} and Alcator C-Mod~\cite{2001_Bombard_etal}. It has been observed that the presence of the narrow feature correlates with low-collisionality in the SOL, carrying substantial power fractions~\cite{2017_Nespoli_etal}. Out of machines, only COMPASS~\cite{2024_Hecko_etal} and ST40~\cite{2024_Zhang_etal} have observed narrow features with near SOL decay lengths on the order of the ion total Larmor radius for diverted discharges. On COMPASS, the extremely narrow heat flux decay lengths were measured in ELMy H-mode discharges both at the divertor target with Langmuir probes and upstream using a Thomson scattering system. The measured narrow footprint decay lengths were a factor of 2~--~3 smaller than what was predicted from scalings~\cite{2013_Eich_etal}, falling below the ion Larmor radius and was considered to be carried by electrons to the target. Further experimental study into the origin of the observed narrow heat flux footprint was not pursued due to COMPASS engineering upgrades. 

ST40 is left with the unique opportunity to experimentally investigate the occurrence of extremely narrow heat flux footprints in diverted H-mode plasmas to inform designs of reactor-relevant devices as the heat flux carried by the narrow feature would be intolerable for long durations in diverted discharges and have implications on detachment access. This article highlights outboard divertor diagnostics in ST40 with specific focus on the Langmuir probe system providing some insight to enhanced heat flux in the near scrape-off layer profile due to large electron currents into the divertor target.

\section{Divertor diagnostics}
\label{sec:Exp_setup}

\begin{figure}[t]
    \centering
    \begin{subfigure}{0.49\columnwidth}
        \centering
        \includegraphics[height=1.29\linewidth]{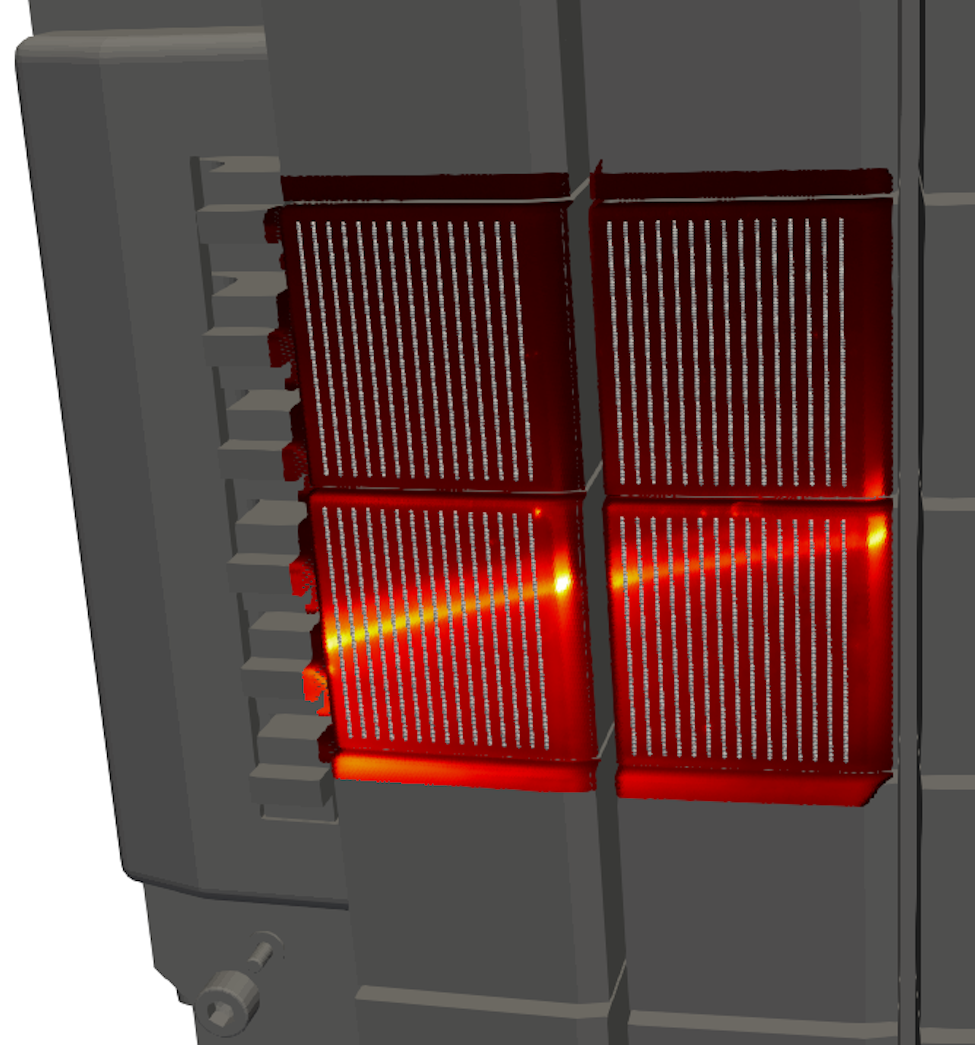}
        \caption{}
        \label{subfig:Paraview_chords_temps_carrier}
    \end{subfigure}
    \hfill
    \begin{subfigure}{0.49\columnwidth}
        \centering
        \includesvg[height=1.3\linewidth]{Figures/ST40_cross_section_13599_192ms.svg}
        \caption{}
        \label{subfig:13599_reconstruction}
    \end{subfigure}
    \caption{IR camera data (gradient coloring) from pulse \#13599 at t~$\approx$~191~ms is mapped to outboard divertor CAD model with chords used by FAHF/IRRITANT for heat flux analysis depicted by white lines. Langmuir probes are placed poloidally along the divertor target flush with the fish-scaled tile geometry. Equilibrium construction (right) shows the last closed flux surface (LCFS) and IR field of view (FOV).}
    \label{fig:CADimage_reconstruction}
\end{figure}

The ST40 tokamak has a major radius of $R$~=~0.4~--~0.5~m, minor radius $a$~=~0.2~m~--~0.25~m and can operate with on-axis magnetic fields up to 2.1~T~\cite{2026_Asunta_etal, 2024_McNamara_etal}. Figure~\ref{fig:CADimage_reconstruction} shows the upper outer divertor with Langmuir probes (LPs) and a temperature profile from an infrared (IR) camera as well as a representative cross-section of ST40 in the $R-Z$ plane with an equilibrium reconstruction for a typical upper single null discharge. Here the ion $\nabla B$ drift is directed upwards. The camera data presented here exhibits a narrow strike-point from ST40 pulse \#13599 at $t\approx$~191~ms with six Langmuir probes located immediately to the left of the mapped data each spaced $\approx$~15~mm.

\subsection{Infrared thermography system and analysis pipeline}
\label{subsec:Infrared_sysdesc}

Divertor temperature profiles are recorded with an IR camera (FLIR X6903sc) operated nominally at a frame rate of 1~kHz calibrated in the range of 20--220$^\circ$~C using a black-body radiation source. Temperature profiles on the divertor tiles are used as boundary conditions for a 2-D grid in a thermographic inversion analysis code, Functional Analysis of Heat Flux (FAHF) for estimating deposited surface heat flux into the plasma facing components~\cite{2025_Moscheni_etal}. In brief, FAHF solves the 2-D heat conduction equation 

\begin{equation}
    \label{eq:2Dheatcond}
    \rho c_p \frac{\partial T}{\partial t} = \nabla \cdot (k \nabla T)
\end{equation}

\noindent where $\rho$ is the mass density, $c_p$ is the specific heat capacity at constant pressure, $T$ is surface temperature along the chord, $t$ is time and $k$ is the material thermal conductivity. Heat flux into the material is then estimated from 

\begin{equation}
    \label{eq:heatfluxeq}
    q_{\perp} = -k \nabla T \cdot \hat{n}
\end{equation}

\noindent where $\hat{n}$ is the surface normal vector of the tile along the chord. Resolution of the 2-D grid is chosen such that the computational algorithm maintains numerical stability~\cite{2025_Moscheni_etal}. 

The calculated surface heat fluxes from FAHF are used as inputs to the InfRaRed InvestigaTive ANlysis Toolchain (IRRITANT) for mapping parallel heat flux profiles to the outboard magnetic mid-plane (OMP) to estimate heat flux decay lengths, accounting for the low-degree of toroidal axisymmetry~\cite{2024_Marsden_etal}. The profiles is fit to different functions such as a single exponential function (so-called ``Eich'' fit)~\cite{2013_Eich_etal} and a double exponential function with the latter giving two $\lambda_q$ values for the near SOL close to the strike point, and for the far SOL. 

\subsection{Divertor Langmuir probes}
\label{subsec:Langmuirprobe_desc}

An array of six Langmuir probes are located besides the divertor tiles in both the upper and lower outboard divertor regions of ST40. The top surface of the probes are  flush to the adjacent fish-scaled divertor tiles with a collection area of 4~mm~$\cross$~10~mm. The Langmuir probes are operated as single-tip probes where a voltage sweep is applied to produce a current--voltage ($I$--$V$) trace. Nominally, each probe is swept with a sinusoidal waveform in the range of -110~--~+50~V (with respect to ST40 ground) at 4~kHz though any arbitrary waveform or fixed potential may be applied in the range of -110~--~+100~V. Data is acquired at 500~kHz but parameter estimates are only provided every half sweep-period after fitting to the function

\begin{equation}
    \label{eq:threeparamfit}
    I  = I_{sat} \left(1 -  exp \left (\frac{V_b - V_f}{T_e}\right ) \right)
\end{equation}

\noindent where $V_b$ is the probe bias voltage, and the floating potential $V_f$, electron temperature $T_e$ and ion saturation current $I_{sat}$ are found from the fit. The LP measurements presented herein were only produced using a three-parameter fit as overestimation of the electron temperature was not considered to be a major concern because the magnetic field intersection angle is large, $>5^\circ$, and the rectangular LP geometry is less susceptible to non-saturation effects as found with probes used on Alcator C-mod~\cite{2019_Podolnik_etal, 2017_Kuang_etal}. Ongoing engineering upgrades to ST40 will extend the LP voltage range to -200--100~V to better accommodate use of a four-parameter fit which accounts for non-saturation of the ion current~\cite{2019_Podolnik_etal, 1990_Matthews_etal, 1995_Gunn_etal}.

After fitting to raw data, electron density at the sheath edge is calculated by

\begin{equation}
    \label{eq:LP_density}
    n_{e,se} = \frac{|I_{sat}|}{eA_p\mathrm{sin}(\alpha)\sqrt{2eT_e/m_i}}
\end{equation}

\noindent where $A_p$ is the probe collection area, $\alpha$ is the total angle between the magnetic field line and the probe surface, $e$ is the elementary charge constant and $m_i$ is the mass of a deuterium ion. The form of Eq.~(\ref{eq:LP_density}) assumes $n_i = n_e$ and $T_e = T_i$. The magnetic equilibrium reconstruction code GSFit~\cite{GSFIT} is used to estimate $\alpha$ for calculating the effective probe collection area $A_{eff} = A_p \mathrm{sin}(\alpha)$ in time and for mapping probe data to the normalized poloidial flux coordinate $\Psi_N$.

Parallel heat flux to the target is calculated by

\begin{equation}
    \label{eq:LP_perp_heatflux}
    q_{\parallel} = \gamma_{sheath} T_e \frac{|I_{sat}|}{A_p \mathrm{sin}(\alpha)}
\end{equation}

\noindent where $\gamma_{sheath}$ is the sheath heat transmission coefficient calculated by

 \begin{equation}
     \label{eq:sheathtransBrunner}
     \begin{split}
              \gamma_{sheath} = 2.5\frac{T_i}{T_e} + 2 \left(1 - \frac{J_{gnd}}{J_{sat}}\right)\\ + \mathrm{ln} \left[ \frac{1}{\sqrt{2 \pi \frac{m_e}{m_i}\left(1+ \frac{T_i}{T_e}\right)}\left(1 - \frac{J_{gnd}}{J_{sat}}\right)}\right]
     \end{split}
 \end{equation}

 \noindent where $T_i$ is the ion temperature, $m_e$ is the electron mass, $J_{sat}$ is the ion saturation current density ($I_{sat}/A_{eff}$). he current density to the grounded target surface, $J_{gnd}$, is found from Langmuir probe data by solving Eq.~(\ref{eq:threeparamfit}) for $V_b~=~0$ yielding $q_\parallel$ at each time step in a ST40 discharge. This form applies for an electrically biased surface with respect to the plasma, in this case a grounded surface ($V = 0$), and is written with the assumptions that deuterium is the only ion species and secondary electron emission effects are negligible~\cite{StangebyTextbook, 2012_Brunner_LaBombard}. Again $T_i = T_e$ is assumed in calculating $\gamma_{sheath}$ here.

Errors in the non-linear least squares fitting of Eq.~(\ref{eq:threeparamfit}) are propagated through all subsequent parameters following a first-order Taylor-series expansion approach~\cite{Numerical_methods_book} at each time step. Individual LP electronic circuits follow those used at TCV~\cite{2019_De_Oliveira_et_al} and, depending on the pulse, may be overloaded when the strike point is in the vicinity of the probe yielding unusable data. All data with fits possessing $R^2~<~0.9$ or those identified to be overloaded are omitted from the analysis output.

\begin{figure}[t]
    \centering
    \includesvg[width=\linewidth]{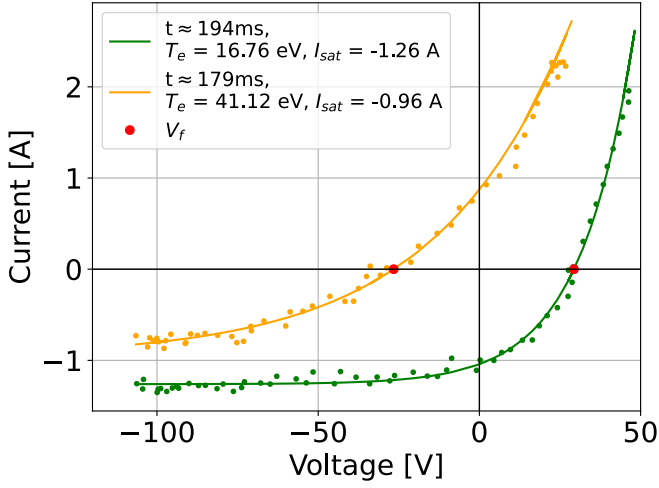}
    \caption{Example of fitting done to raw data for ST40 pulse 13599, both fits possess $R^2>$.96.}
    \label{fig:exampledata}
\end{figure}

\begin{figure}[t]
    \centering
    \begin{subfigure}{0.49\textwidth}
        \centering
        \includesvg[width=\textwidth]{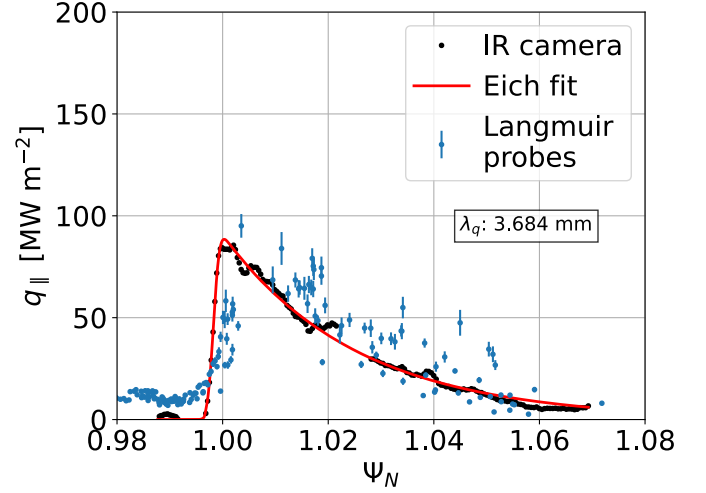}
        \caption{\#13548 (L-mode)}
        \label{subfig:13548_qbvsPsiN}
    \end{subfigure}
    \hfill
    \begin{subfigure}{0.49\textwidth}
        \centering
        \includesvg[width=\textwidth]{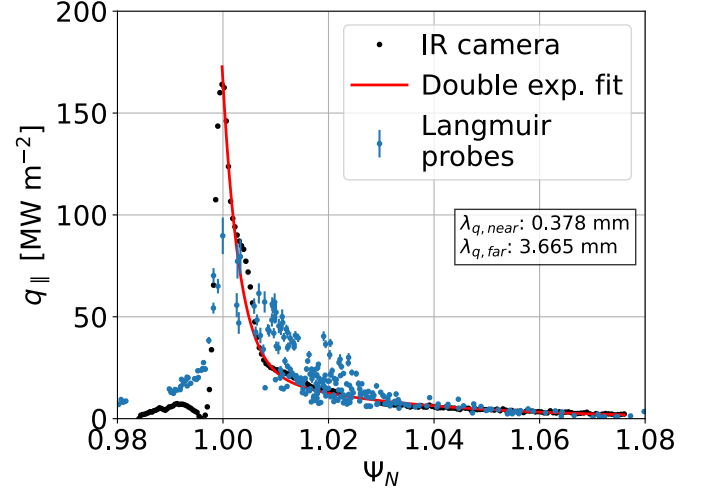}
        \caption{\#13599 (H-mode)}
        \label{subfig:13599_qbvsPsiN}
    \end{subfigure}
    \caption{Target parallel heat flux from Langmuir probe and IR camera measurements. Fit profiles shown are only constructed using the IR camera data for a singular chord on the outboard divertor. Decay lengths shown correspond to values at OMP.}
    \label{fig:QparallelvsPsi}
\end{figure}

Examples of raw LP data with fits from Eq.~(\ref{eq:threeparamfit}) are shown in Fig.~\ref{fig:exampledata} for two time-steps from the same probe in ST40 pulse 13599. Equation~(\ref{eq:threeparamfit}) adequately fits to the data for both probe sweeps with high $R^2$ values though the ion saturation region of the orange colored data is not clear from available measurements. With these data, good agreement can be found when comparing LP measurements of $q_{\parallel}$ to the IR thermography system along the target evident in Fig.~\ref{fig:QparallelvsPsi} which shows results for two ST40 pulses. Some minor disagreements between the two systems could be attributed to the fact that the probes inherently provide spatially averaged data giving slightly higher values and different slopes of $q_\parallel$. Despite the assumptions made in the LP analysis noted above, the two diagnostic systems are in satisfactory agreement.

\section{Comparison of L-mode and H-mode pulse pair}
\label{sec:PulsePair}
\begin{figure}[t]
    \centering
    \includesvg[width=\linewidth]{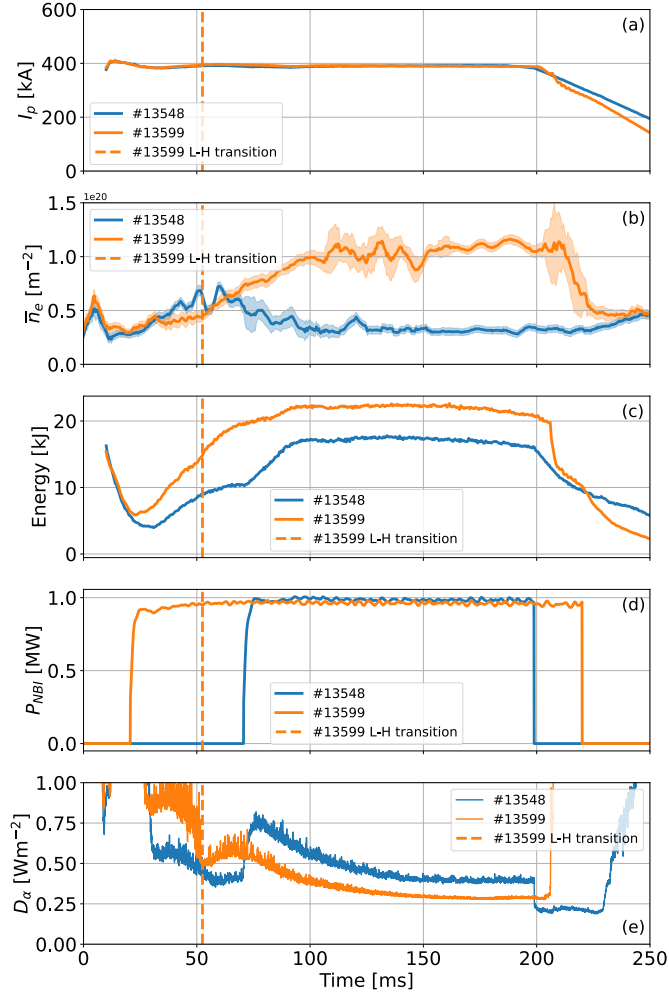}
    \caption{(a) Plasma current, (b) smoothed line-integrated density, (c) stored energy, (d) neutral beam input power and (e) D-$\alpha$ emission from two ST40 pulses with $B_t$~=~1.8~T.}
    \label{fig:pulse_overview}
\end{figure}

Two similar plasma pulses from the ST40 database are presented each with a plasma current of 400~kA and toroidal magnetic field of 1.8~T. An overview of both pulses for core line-integrated densities, stored energy, neutral beam heating and $D_\alpha$ emission is shown in Fig.~\ref{fig:pulse_overview}. Pulse \#13599 enters H-mode at 52~ms indicated by the vertical dashed line and persists beyond 200~ms. Langmuir probe and IR camera data presented herein is from time windows of $t=$~110--130~ms and $t=$~145--200~ms for pulses \#13548 and \#13599 respectively such that the outer strike point was in-view of the IR camera with and possessed minor fluctuations. Presented LP and IR camera data is shown using the normalized poloidal flux coordinate, $\psi_N$, though profiles are shifted such that the peak heat flux is aligned with $\psi_N$~=~1. This shift corresponds to a distance of 2.3~mm and 1.4~mm at the OMP for pulse \#13548 and \#13599 respectively within the uncertainty of the equilibrium reconstruction. There was also no active control feedback of strike point location for the pulses presented but typically the strike point sweep outs radially along the outer target during a pulse due to induced current in the divertor plates enabling the LP array to measure complete profiles.

\begin{figure}[t]
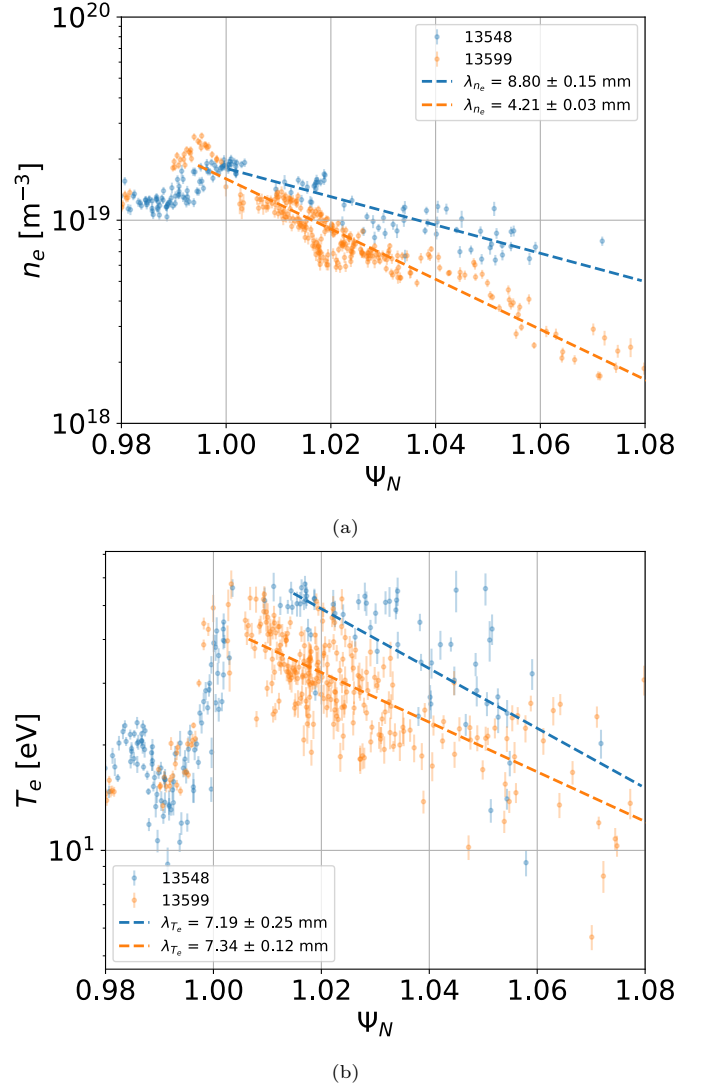

    \centering
    \begin{subfigure}{0.49\textwidth}
        \centering
        \includesvg[width=\textwidth]{Figures/Ne_vs_psiN_combined.svg}
        \caption{}
        \label{subfig:NevsPsiN}
    \end{subfigure}
    \hfill
    \begin{subfigure}{0.49\textwidth}
        \centering
        \includesvg[width=\textwidth]{Figures/Te_vs_psiN_combined.svg}
        \caption{}
        \label{subfig:TevsPsiN}
    \end{subfigure}
    \caption{Electron density and temperature measurements in outboard divertor region. Data fit to single exponential function with decay lengths presented at OMP.}
    \label{fig:NeTevsPsi}
\end{figure}

Electron density and temperature profiles at the outer divertor target are shown in Fig.~\ref{fig:NeTevsPsi} with data fit to single exponential functions in the form of

\begin{equation}
    \label{eq:density_lambda}
    n_e(r-r_{0}) = n_{e,0}\mathrm{exp}(-(r-r_{0})/\lambda_{n_e})
\end{equation}

\begin{equation}
\label{eq:temperature_lambda}
    T_e(r-r_{0}) = T_{e,0}\mathrm{exp}(-(r-r_{0})/\lambda_{T_e})
\end{equation}

\noindent where $r-r_{0}$ is the distance along the target from the strike point location. Density and temperature decay lengths, $\lambda_{n_e}$ and $\lambda_{T_e}$, are shown after mapping the data to the OMP. Electron density profiles are narrower in H-mode compared to L-mode given by the smaller $\lambda_{n_e}$, a behavior consistent with the observed heat flux profiles in Fig.~\ref{fig:QparallelvsPsi}. The temperature profile is however slightly broader in H-mode but due to the noisy temperature data. The H-mode heat flux profile is much narrower than the density and temperature profiles in Fig.~\ref{fig:NeTevsPsi}, suggesting the narrow heat flux profile is not dominated by sharper density or temperature gradients near the strike point.

\begin{figure}[t]
    \centering
    \begin{subfigure}{0.49\textwidth}
        \centering
        \includesvg[width=\textwidth]{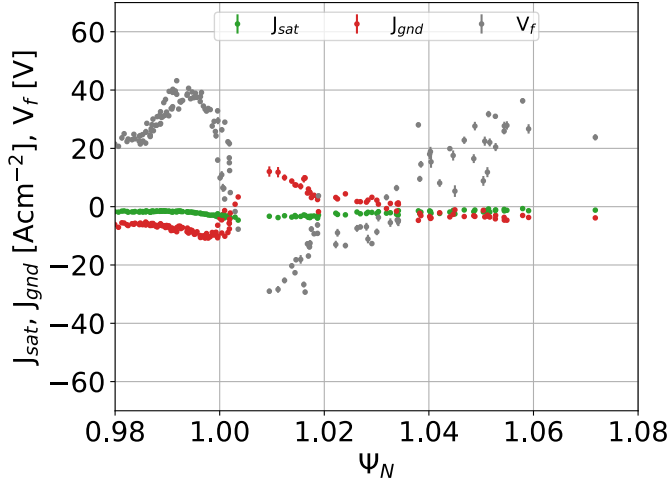}
        \caption{\#13548 (L-mode)}
        \label{subfig:13548_JsatJgndVfvsPsi}
    \end{subfigure}
    \hfill
    \begin{subfigure}{0.49\textwidth}
        \centering
        \includesvg[width=\textwidth]{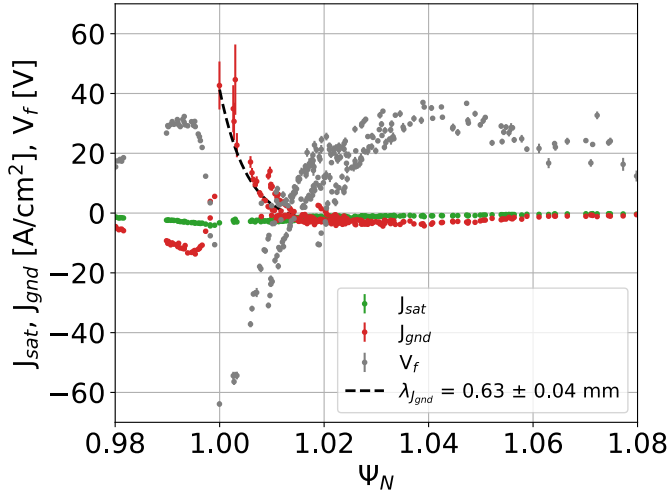}
        \caption{\#13599 (H-mode)}
        \label{subfig:13599_JsatJgndVfvsPsiN}
    \end{subfigure}
    \caption{Target profiles of $J_{sat}$, $J_{gnd}$ and $V_f$ measured by divertor Langmuir probes. A fit similar to Eqs.~(\ref{eq:density_lambda}) and~(\ref{eq:temperature_lambda}) is applied to $J_{gnd}$ with the length scale shown corresponding to the profile at OMP.}
    \label{fig:JsatJgndVfvsPsi}
\end{figure} 

Profiles of $J_{sat}$, $J_{gnd}$ and $V_f$  are shown in Fig.~\ref{fig:JsatJgndVfvsPsi}. In both pulses, $V_f$ drops from the private-flux region to the common-flux region coinciding with an inverse behavior for $J_{gnd}$. Ion collection by the Langmuir probes is defined as negative current indicating the ground current density is carried mostly by electrons near $\Psi_N~=~1$ in both cases but much greater in H-mode pulse \#13599. A sharp drop in the H-mode $V_f$ profile is similar to one observed on TCV~\cite{2017_Nespoli_etal} in the near SOL of a limited plasma discharge where the existence of this feature was found to be linked to low collisionality.

An exponential fit was applied to $J_{gnd}$ for the H-mode pulse and a length scale $\lambda_{J_{gnd}}$, is used here to illustrate the decay length scale of the electron current channel. The $\lambda_{J_{gnd}}$ profile in the H-mode discharge is close to $\lambda_{q,near}$ attributed to the narrow feature in the SOL. Niether electron density nor temperature profiles exhibit a profile akin to the near SOL narrow feature. This strongly suggests the narrow near SOL heat flux is carried by an enhanced electron current to the target.

\begin{figure}[t]
    \centering
    \includesvg[width=\linewidth]{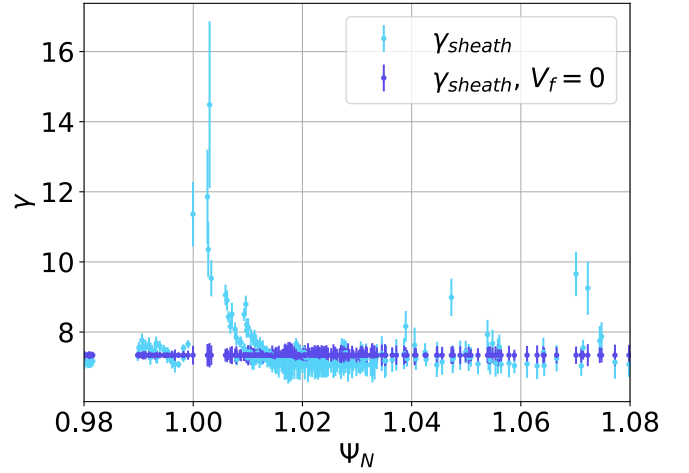}
    \caption{Sheath transmission coefficient for H-mode pulse \#13599 at the divertor target with and without enforcing zero ground current density.}
    \label{fig:sheathtrans}
\end{figure}

 The sheath transmission coefficient inferred from the LPs are shown in Fig.~\ref{fig:sheathtrans} only for the H-mode pulse \#13599 with and without enforcing $V_f=0$ or zero ground current density. There is a prominent peak near the strike point in $\gamma_{sheath}$ in agreement with what is observed in the $J_{gnd}$ profile and the IR camera heat flux profiles. When $V_f=0$ is enforced, $\gamma_{sheath}$ falls to the standard assumed range of 7--8 as expected~\cite{StangebyTextbook}. It is clear the inclusion of the $J_{gnd}$ terms in Eq.~(\ref{eq:sheathtransBrunner}) is crucial in capturing the narrow heat flux footprint and further corroborates that the near SOL narrow feature is related to enhanced electron current carried into the target. The source of this enhanced electron current is unknown but may be linked to the fish-scaled divertor tile geometry itself where ions in their gyro-orbits are lost to the tiles creating a positive potential hill thereby accelerating electrons to the target~\cite{Chang_Fish_scale}.

\section{Summary and future work}
\label{sec:conclusions}
The ST40 tokamak is one of few devices to-date to observe extremely narrow heat flux profiles on the order of the ion Larmor radius which deviates from established scaling laws. An array of Langmuir probes and an IR camera thermography system were used in tandem to measure parallel heat flux profiles in the upper outboard divertor region in L-mode and ELM-free H-mode discharges with good agreement between the two diagnostics. Langmuir probe measurements near the outer strike point revealed an increase in ground current density primarily carried by electrons with an exponential length scale similar to the narrow SOL heat flux decay length inferred from IR camera measurements. Inclusion of the ground current density in the sheath transmission coefficient further verifies the correlation with electron-carried current and the presence of a narrow heat flux footprint. Future ST40 experimental campaigns will leverage a new lower divertor IR thermography system and additional inboard/outboard divertor Langmuir probes for SOL width studies in lower single-null and double-null plasmas. Additional Langmuir probes will aid in determining current flow in ST40 to help understand the origin of the enhanced electron current as well as power fractions carried in the near and far SOL.

\section*{Acknowledgments} 
This work was supported in part by the UK Department for Energy Security and Net Zero and the Science and Technology Facilities Council (ST/LEAPS/0925) and U.S. Department of Energy (DE-SC0025884 and NFE-24-10427).
\appendix

\bibliography{biblio.bib}

\end{document}